\pdfoutput=1
\documentclass[preprint,preprintnumbers,amsmath,amssymb,floatfix,endfloats*]{revtex4}

\usepackage{graphicx}
\usepackage{dcolumn}
\usepackage{bm}
\usepackage{amsfonts}

\DeclareMathAlphabet{\mathsfsl}{OT1}{cmr}{bx}{it}
\begin{document}
%
\title{Structural relaxation and delayed yielding in cyclically sheared Cu-Zr metallic glasses}
\author{Nikolai V. Priezjev$^{1,2}$}
\affiliation{$^{1}$Department of Civil and Environmental
Engineering, Howard University, Washington, D.C. 20059}
\affiliation{$^{2}$Department of Mechanical and Materials
Engineering, Wright State University, Dayton, OH 45435}
\date{\today}
\begin{abstract}

The yielding transition, structural relaxation, and mechanical
properties of metallic glasses subjected to repeated loading are
examined using molecular dynamics simulations. We consider a
poorly-annealed Cu-Zr amorphous alloy periodically deformed in a
wide range of strain amplitudes at room temperature. It is found
that low-amplitude cyclic loading leads to a logarithmic decay of
the potential energy, and lower energy states are attained when the
strain amplitude approaches a critical point from below. Moreover,
the potential energy after several thousand loading cycles is a
linear function of the peak value of the stress overshoot during
startup continuous shear deformation of the annealed sample. We show
that the process of structural relaxation involves collective,
irreversible rearrangements of groups of atoms whose spatial extent
is most pronounced at the initial stage of loading and higher strain
amplitudes. At the critical amplitude, the glass becomes
mechanically annealed for a number of transient cycles and then
yields via formation of a shear band. The yielding transition is
clearly marked by abrupt changes in the potential energy, storage
modulus, and fraction of atoms with large nonaffine displacements.

\vskip 0.5in


\end{abstract}

\maketitle

\section{Introduction}

Establishing structure-property-performance relations for bulk
metallic glasses is important for various structural and functional
applications~\cite{Gao22,Jiang21}.  Owing to their amorphous atomic
structure, metallic glasses offer a number of unique properties,
such as high strength, large elastic limit, as well as superior wear
and corrosion resistance~\cite{Egami13,Qiao19}. An outstanding
challenge that limits their widespread use is that sufficiently well
annealed glasses are prone to brittle failure via formation of
nanoscale shear bands~\cite{Hufnagel16}. In addition, metallic
glasses subjected cyclic variations in stresses or strains exhibit
relatively low fatigue limit and fatigue life, which can be affected
by the sample size, chemical composition, cycling frequency, and
surface conditions~\cite{Liaw18,Menzel06,Sha2020}. To enhance their
ductility, metallic glasses can be rejuvenated using several
thermo-mechanical processing methods, including high-pressure
torsion, ion irradiation, thermal cycling, cold rolling, and
elastostatic loading~\cite{Greer16,Ketov15}. Alternatively, an
atomic structure can be simply reset by heating above the glass
transition temperature and subsequent rapid cooling to the glass
state or a supercooled glass former can be frozen under applied
stress~\cite{Mota21,PriezMETAL21}. In recent years, laser-based
additive manufacturing techniques were also introduced to fabricate
large-scale complex structures and patient-specific implants for
biomedical applications~\cite{Reed23}. Despite these advances,
however, development of novel processing methods to rejuvenate
metallic glasses and improve their mechanical properties remains a
challenging task.

\vskip 0.05in

In the last decade, a number of molecular dynamics (MD) simulation
studies have investigated the yielding behavior and structural
relaxation in amorphous solids subjected to oscillatory
deformation~\cite{Priezjev13,Reichhardt13,Sastry13,IdoNature15,
Shi15,GaoNano15,Priezjev16,Kawasaki16,Priezjev16a,Sastry17,
Priezjev17,Hecke17,Priezjev18,Priezjev18a,Sastry19band,
NVP18strload,PriezSHALT19,Priez20ba,KawBer20,NVP20altY,
Priez20delay,BhaSastry21,Priez21var,PriezCMS21,Peng22,Sastry22,
PriezJNCS22,Barrat23,PriezCMS23,Karmakar23,PriezMet23,Egami24,
Tang24}.  Notably, it was demonstrated that in the athermal limit,
binary Lennard-Jones (LJ) glasses under low-amplitude loading
gradually evolve into periodic limit cycles with exactly reversible
trajectories of atoms~\cite{Reichhardt13,IdoNature15}, and the
potential energy of such states is lower for better annealed glasses
and higher strain amplitudes~\cite{Sastry13,Sastry17}.   By
contrast, periodic shear deformation in combination with thermal
noise facilitate collective, irreversible rearrangements of atoms
and prolonged structural relaxation~\cite{Sastry22,PriezJNCS22}. In
addition, lower energy states can be accessed when cyclic shear is
periodically alternated along two or three mutually perpendicular
planes~\cite{PriezSHALT19,Karmakar23}.  Moreover, the range of
energy states attainable in thermal glasses during low-amplitude
loading can be extended by increasing strain amplitude slightly
above a critical point every few cycles~\cite{Priez21var}.
Furthermore, when the loading amplitude exceeds a critical value,
amorphous alloys undergo a yielding transition after a number of
transient cycles~\cite{GaoNano15,Sastry17,Priezjev17,Priezjev18a,
Sastry19band,Priez20ba,NVP20altY,Priez20delay,PriezCMS21,
PriezJNCS22,PriezCMS23,PriezMet23}. Interestingly, the critical
strain amplitude in athermal systems might depend on the glass
stability~\cite{KawBer20,BhaSastry21}, whereas at about half $T_g$,
the yielding transition is delayed in better annealed glasses but
the critical strain amplitude remains unchanged~\cite{PriezJNCS22}.
It was recently shown that well annealed binary glasses fail via
shear band formation, and number of cycles to reach the yielding
transition is well described by the power-law function of the
difference between the strain amplitude and its critical
value~\cite{PriezCMS23,PriezMet23}. However, the role of loading
conditions and preparation history on the critical behavior of
metallic glasses is not yet fully understood.

\vskip 0.05in

In this paper, the yielding behavior and structural relaxation in a
Cu-Zr metallic glass under periodic shear deformation are studied
using molecular dynamics simulations. We consider the binary glass
that is first rapidly cooled across the glass transition to room
temperature and then cyclically loaded in a wide range of strain
amplitudes around a critical value. It will be shown that
low-amplitude loading leads to a logarithmic decay of the potential
energy during thousands of cycles. When the loading amplitude
increases toward the critical value, the average size of plastically
deformed domains becomes larger and the glass is relocated to lower
energy states. On the contrary, we find that the yielding transition
at the critical strain amplitude is marked by abrupt changes in the
storage modulus, potential energy, and the number of atoms with
large nonaffine displacements, which are localized within a narrow
shear band.

\vskip 0.05in

The rest of this paper is structured as follows. The MD simulation
setup, parameter values, as well as cooling and deformation
protocols are described in the next section. The analysis of the
potential energy, shear stress, and nonaffine displacements are
presented in section\,\ref{sec:Results}. A summary of the results is
provided in the last section.

\section{MD simulations}
\label{sec:MD_Model}

In our study, the metallic glass was represented by a binary mixture
of Cu and Zr atoms, which interacted via the embedded atom method
(EAM) potentials~\cite{CuZrEAM09,Ma11}. The total number of atoms in
the $\text{Cu}_{50}\text{Zr}_{50}$ glass is $60\,000$. The
preparation procedure consisted of several steps. First, the binary
mixture was thoroughly equilibrated in a periodic box at the
temperature of $2000\,\text{K}$ and zero pressure. This temperature
is well above the glass transition temperature $T_g\approx
675\,\text{K}$. We followed the cooling protocol by Fan and
Ma~\cite{FanMa21}, where the Cu-Zr system is initially cooled to
$1500\,\text{K}$ at the rate of $10^{13}\,\text{K/s}$ and then to
$300\,\text{K}$ at $10^{12}\,\text{K/s}$. As a result, the effective
cooling rate was $10^{12}\,\text{K/s}$. The sample was prepared
using the Nos\'{e}-Hoover thermostat, zero applied pressure, and
periodic boundary conditions. The equations of motion were
integrated with the time step $\triangle t =
1.0\,\text{fs}$~\cite{Lammps}.

\vskip 0.05in


The periodic shear deformation was applied along the $xz$ plane as
follows:
\begin{equation}
\gamma_{xz}(t)=\gamma_0\,\text{sin}(2\pi t/T ),
\label{Eq:shear}
\end{equation}
where $\gamma_0$ is the strain amplitude and $T$ is the oscillation
period. In our setup, the oscillation period was set to
$1.0\,\text{ns}$ ($10^6$ MD time steps), and the strain amplitude
was varied in the range $0.020 \leqslant \gamma_0 \leqslant 0.061$.
The MD simulations were carried out in the NVT ensemble with the
temperature of $300\,\text{K}$ and the linear size of the periodic
box of $101.8\,\text{\AA}$. A typical simulation of 4000 shear
cycles took about 6000 hours using 40 processors in parallel. Due to
computational limitations, the data for the potential energy, shear
stress, and atomic configurations were collected for only one
independent sample.

\section{Results}
\label{sec:Results}


One of the key factors that strongly affects mechanical properties
of metallic glasses is the rate of cooling during glass
formation~\cite{Greer16}.  In general, glasses obtained by cooling
at a slower rate become more stable and, upon deformation, exhibit a
stress overshoot followed by plastic flow, whereas rapidly cooled
glasses are settled at higher energy states and yield more
smoothly~\cite{Egami13}.  Alternatively, it was recently
demonstrated that poorly annealed LJ glasses can be relocated to
lower energy states via low-amplitude cyclic loading at a finite
temperature below $T_g$~\cite{Sastry22,PriezJNCS22}. However, the
resulting change in the potential energy at given strain amplitude
and temperature as well as the yielding behavior near the critical
strain amplitude for more realistic models of glasses remain to be
determined. In the present study, we considered the
$\text{Cu}_{50}\text{Zr}_{50}$ glass that was first rapidly cooled
to room temperature and then periodically strained for $4000$ shear
cycles over a broad range of strain amplitudes. This relatively
large number of loading cycles was set based on the results of the
previous MD study where a rapidly cooled binary LJ glass (60\,000
atoms) yielded only after about $2500$ cycles at a critical strain
amplitude~\cite{NVP20altY}.

\vskip 0.05in

\begin{figure}[t]
\includegraphics[width=12.0cm,angle=0]{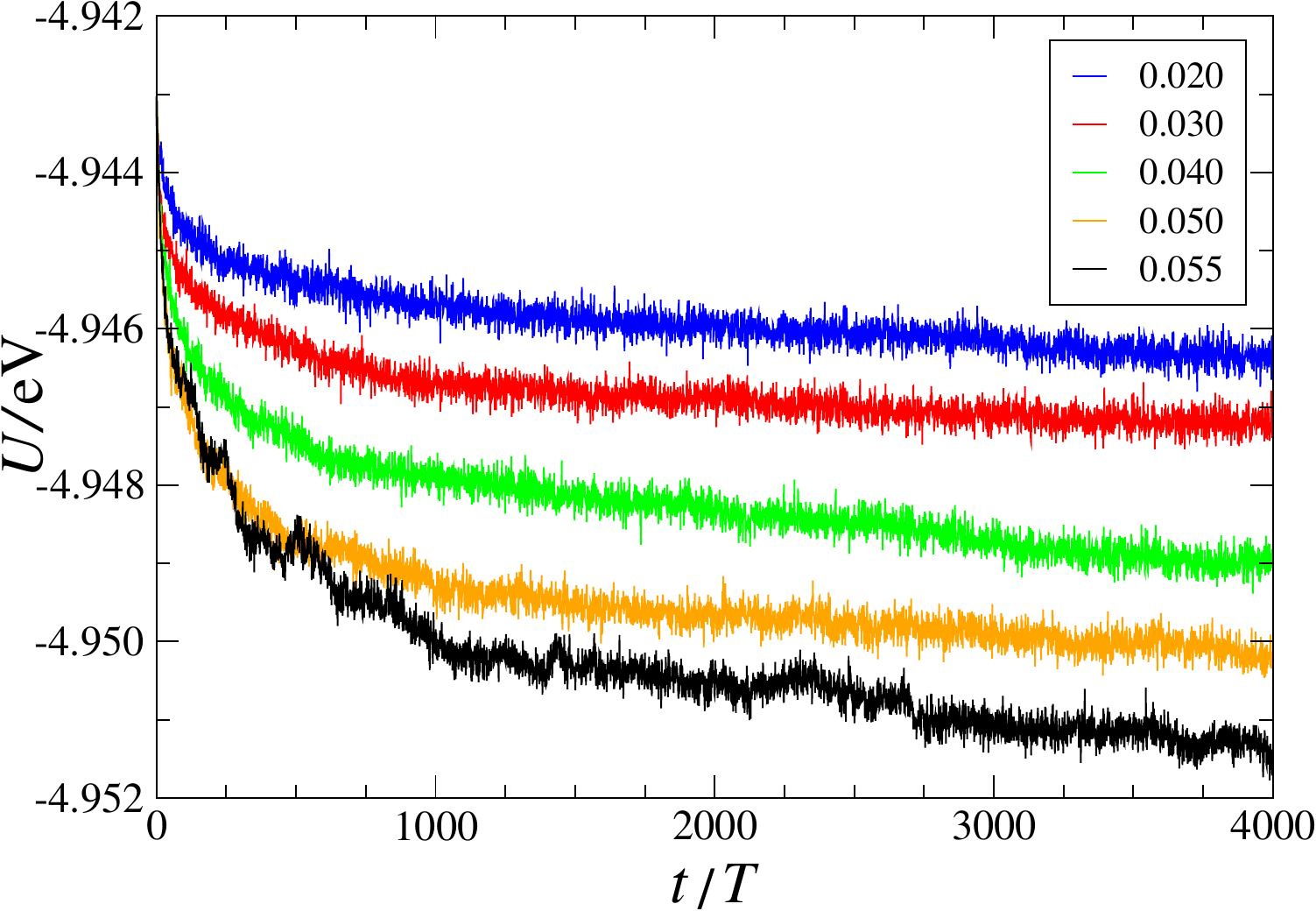}
\caption{The potential energy at the end of each cycle versus cycle
number for the indicated strain amplitudes $\gamma_0$. The period of
oscillation is $T=1.0\,\text{ns}$.}
\label{fig:poten_T300_r10e12_amp_020_055}
\end{figure}
\begin{figure}[t]
\includegraphics[width=12.0cm,angle=0]{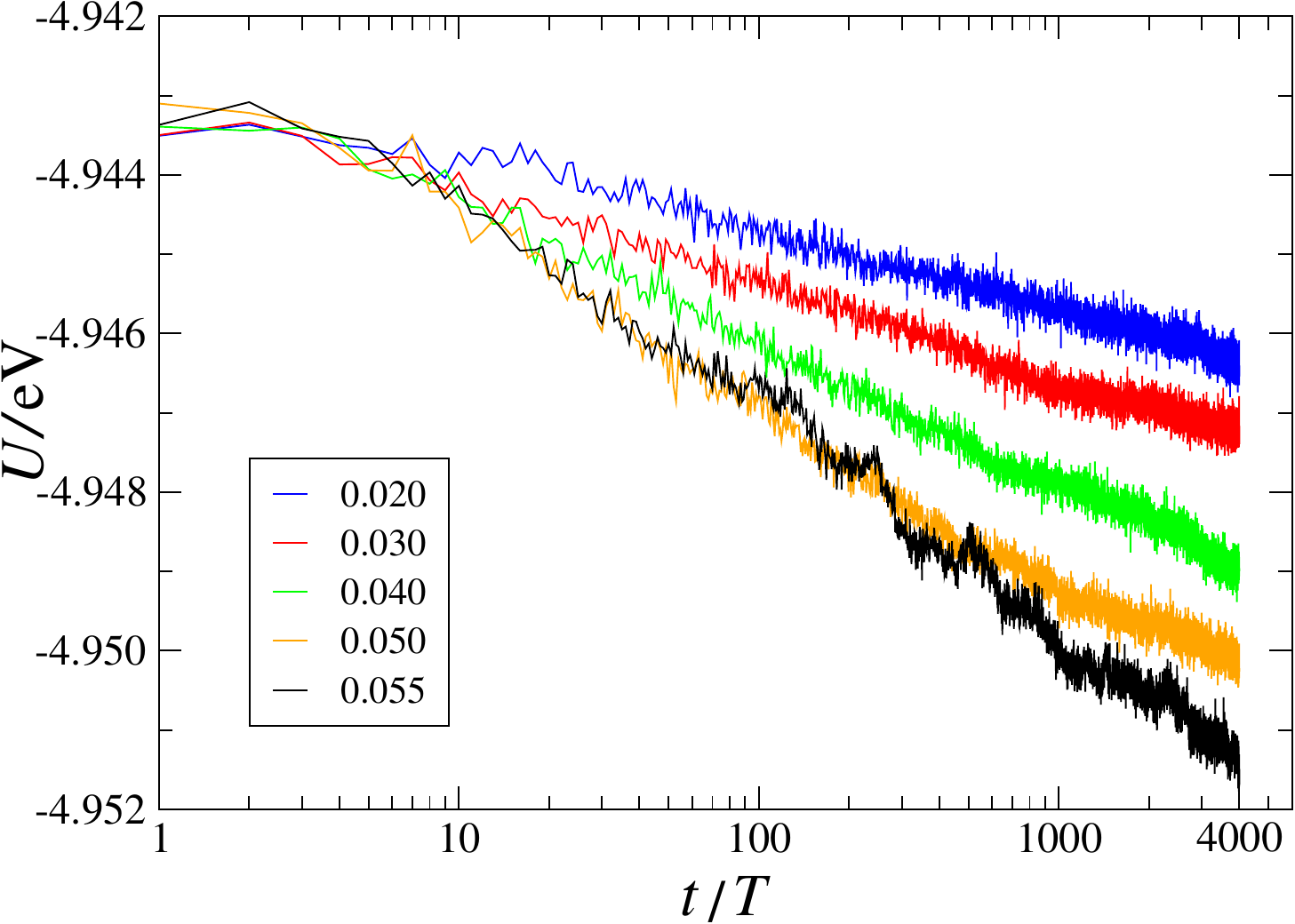}
\caption{A linear-log plot of the same data as in
Fig.\,\ref{fig:poten_T300_r10e12_amp_020_055}.}
\label{fig:poten_T300_r10e12_amp_020_055_log}
\end{figure}

We first plot potential energy minima after each cycle at zero
strain in Fig.\,\ref{fig:poten_T300_r10e12_amp_020_055} for strain
amplitudes $0.020 \leqslant \gamma_0 \leqslant 0.055$. As is
evident, the glass is relocated to progressively lower energy states
upon continued loading, and cycle-to-cycle fluctuations are enhanced
at larger strain amplitudes. Moreover, cyclic loading at higher
strain amplitudes (up to a critical value) allows for rearrangement
of larger clusters of atoms during each cycle, leading, on average,
to more relaxed states. These conclusions are in agreement with the
results of previous MD studies of binary LJ
glasses~\cite{Sastry17,Priezjev18,Sastry22,PriezJNCS22}. In
Fig.\,\ref{fig:poten_T300_r10e12_amp_020_055_log}, the same data for
the potential energy as a function of the cycle number are replotted
on a semi-log scale. It can be clearly observed that for each strain
amplitude, the potential energy closely follows a logarithmic decay
when $t/T\gtrsim 10$, suggesting a possibility of reaching lower
energy states upon further loading. Interestingly, periodic
deformation for 4000 shear cycles at $\gamma_0 = 0.055$ resulted in
the potential energy $U\approx-4.951\,\text{eV}$ (see
Fig.\,\ref{fig:poten_T300_r10e12_amp_020_055_log}), which is nearly
the same as in the case of a better annealed
$\text{Cu}_{50}\text{Zr}_{50}$ glass prepared at a 100 times slower
cooling rate~\cite{PriezMet23}.  In other words, the rapidly cooled
glass ($10^{12}\,\text{K/s}$) was mechanically annealed to the
potential energy level of a more slowly cooled glass
($10^{10}\,\text{K/s}$), which was considered in the previous MD
study~\cite{PriezMet23}.

\vskip 0.05in

\begin{figure}[t]
\includegraphics[width=12.0cm,angle=0]{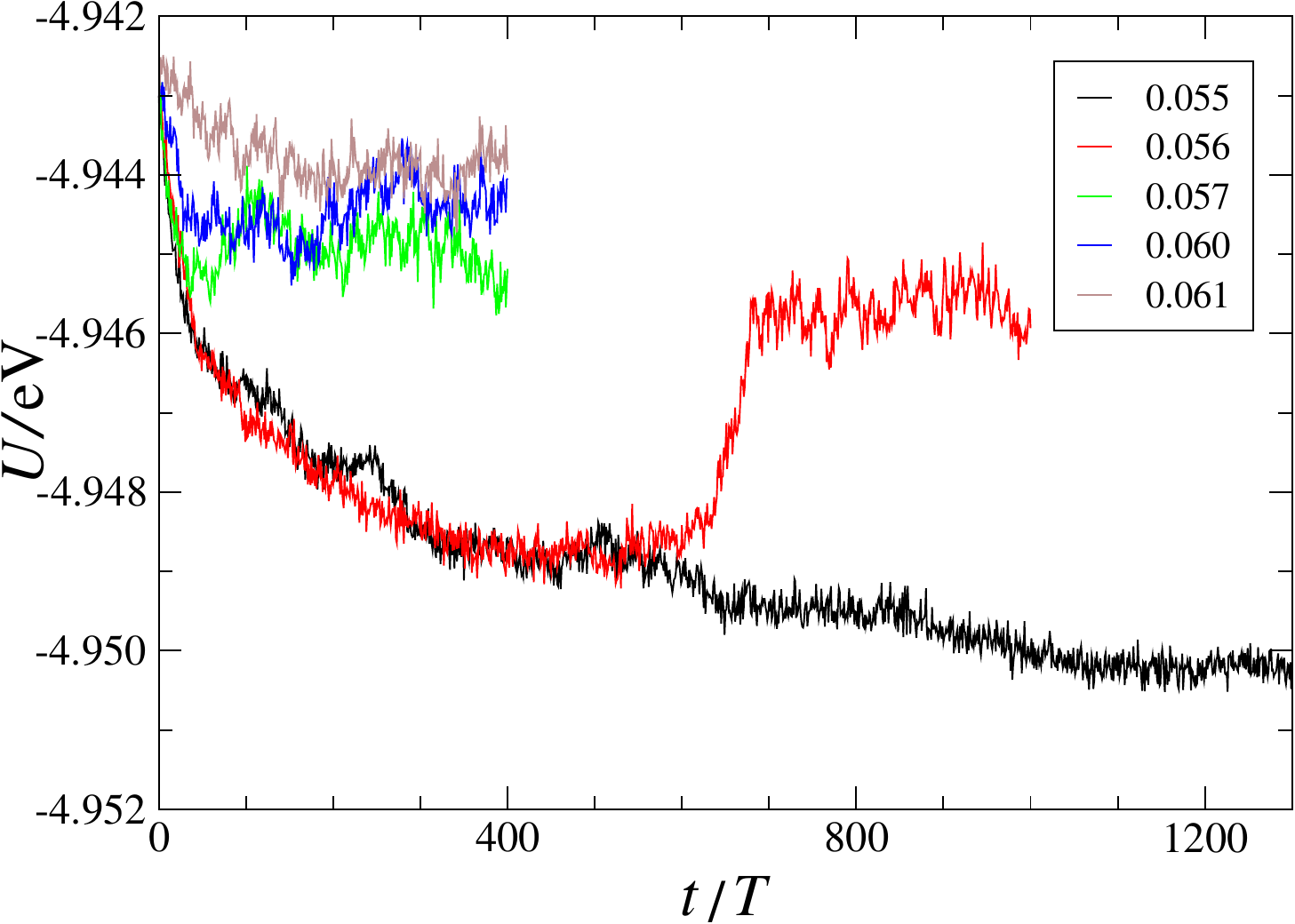}
\caption{The variation of potential energy when strain is zero as a
function of the number of cycles for strain amplitudes $\gamma_0 =
0.055$, $0.056$, $0.057$, $0.060$ and $0.061$. The data for
$\gamma_0 = 0.055$ are the same as in
Fig.\,\ref{fig:poten_T300_r10e12_amp_020_055}. }
\label{fig:poten_T300_r10e12_amp_055_56_57_60_61}
\end{figure}

The variation of the potential energy during cyclic loading at
higher strain amplitudes, $\gamma_0 \geqslant 0.055$, is presented
in Fig.\,\ref{fig:poten_T300_r10e12_amp_055_56_57_60_61}. For
reference, the same data for $\gamma_0=0.055$ as in
Fig.\,\ref{fig:poten_T300_r10e12_amp_020_055} are also included in
Fig.\,\ref{fig:poten_T300_r10e12_amp_055_56_57_60_61} for the first
1300 cycles.  It can be readily seen that, following a number of
transient cycles, the potential energy abruptly increases,
indicating shear band formation at the yielding transition. Note
that except for a relatively large strain amplitude $\gamma_0 =
0.061$, rapidly cooled glass is mechanically annealed for a number
of cycles before yielding. The number of cycles to reach the
yielding transition generally increases upon reducing strain
amplitude toward a critical value. As shown in
Fig.\,\ref{fig:poten_T300_r10e12_amp_055_56_57_60_61}, the maximum
number of transient cycles is about 650 at the critical strain
amplitude $\gamma_0 = 0.056$. The transient behavior can be
rationalized as follows.  When a rapidly cooled glass is initially
strained up to $\gamma_{xz}=\pm \gamma_0$ in Eq.\,(\ref{Eq:shear}),
the maximum stress remains relatively low and plastic deformation is
homogeneously distributed within the sample. Upon further loading,
the glass becomes more stable (lower potential energy at zero
strain), the maximum stress increases, and the formation of a
large-scale plastic event becomes more probable. We finally comment
that a better annealed $\text{Cu}_{50}\text{Zr}_{50}$ glass under
cyclic loading did not yield for 700 cycles at $\gamma_0 = 0.056$,
when the potential energy was
$U\approx-4.951\,\text{eV}$~\cite{PriezMet23}. Instead, the critical
strain amplitude was found to be $\gamma_0 = 0.057$ for similar
loading conditions~\cite{PriezMet23}.

\vskip 0.05in

\begin{figure}[t]
\includegraphics[width=12.0cm,angle=0]{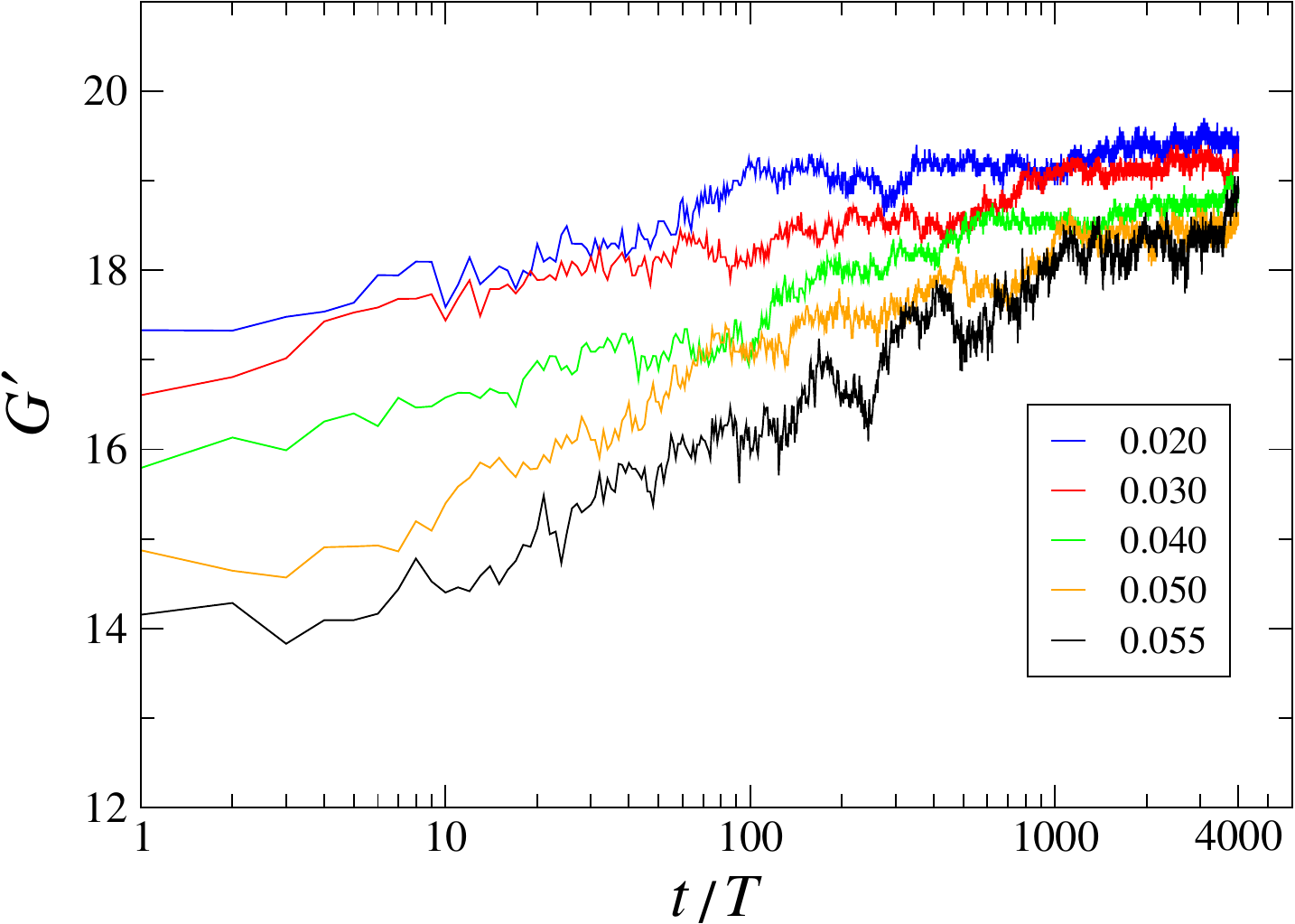}
\caption{The storage modulus $G^{\prime}$ (in units of GPa) as a
function of the cycle number for the indicated values of the strain
amplitude $\gamma_0$. The period of oscillation is
$T=1.0\,\text{ns}$.}
\label{fig:Gp_4000_T300_r10e12_amp_020_055}
\end{figure}

Along with the potential energy, the time dependence of shear
stress, $\sigma_{xz}(t)$, was analyzed for different strain
amplitudes. For each shear cycle, we computed the storage modulus,
$G^{\prime}=\sigma^{max}_{xz}/\gamma_0\,\text{cos}(\delta)$, where
$\delta$ is the phase lag between stress and
strain~\cite{McKinley11}. The variation of the storage modulus is
presented in Fig.\,\ref{fig:Gp_4000_T300_r10e12_amp_020_055} for
strain amplitudes below the critical value, i.e., $\gamma_0
\leqslant 0.055$. It can be seen that for each strain amplitude, the
storage modulus increases roughly logarithmically as a function of
the cycle number, which is consistent with the decay of the
potential energy reported in
Fig.\,\ref{fig:poten_T300_r10e12_amp_020_055_log}. The largest
increase in $G^{\prime}$ during 4000 cycles is found for the strain
amplitude $\gamma_0 = 0.055$, which is just below the critical value
(see Fig.\,\ref{fig:Gp_4000_T300_r10e12_amp_020_055}). Similar to
results for binary LJ glasses~\cite{Priezjev18a}, the storage
modulus is larger for cyclic loading at smaller strain amplitudes,
where deviation from the elastic regime of deformation is reduced.

\vskip 0.05in

\begin{figure}[t]
\includegraphics[width=12.0cm,angle=0]{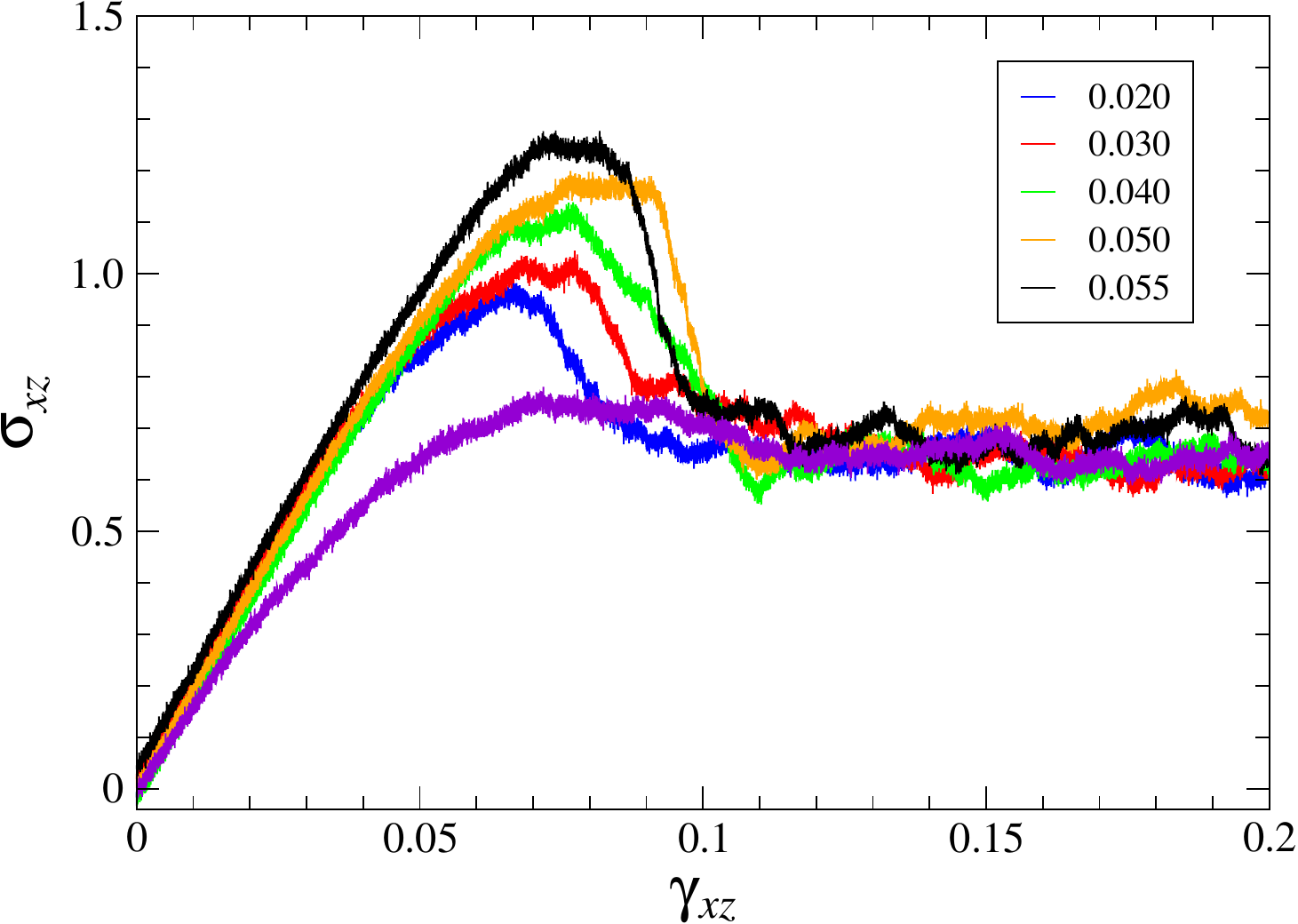}
\caption{Shear stress, $\sigma_{xz}$ (in units of GPa), versus
strain for glasses deformed at a constant strain rate of
$10^{-5}\,\text{ps}^{-1}$ after 4000 cycles at indicated $\gamma_0$.
The lower violet curve is the data before cyclic loading.}
\label{fig:stress_strain_const_amps_upto_055}
\end{figure}

\begin{figure}[t]
\includegraphics[width=12.0cm,angle=0]{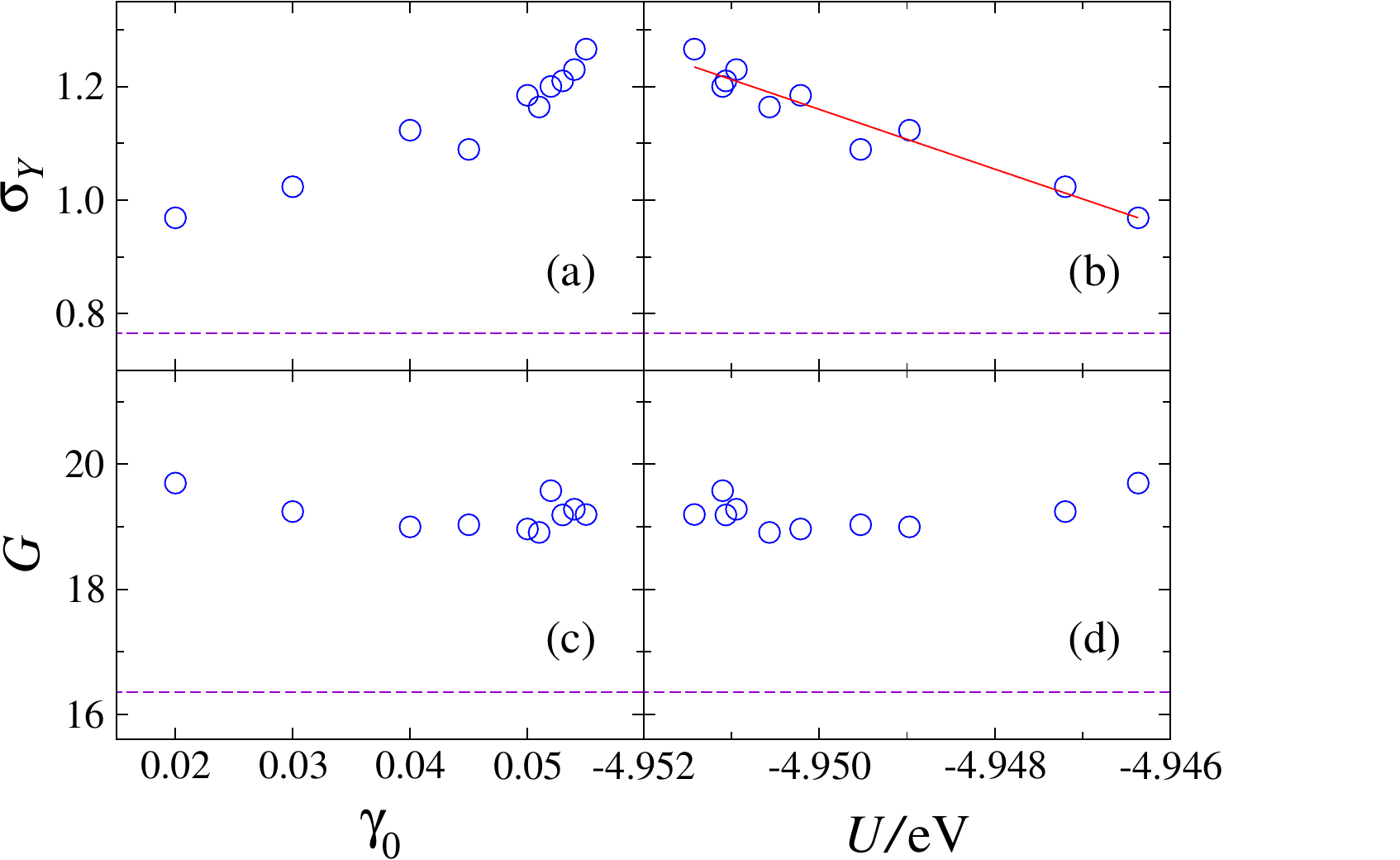}
\caption{The shear modulus, $G$ (in GPa), and the peak value of the
stress overshoot, $\sigma_Y$ (in GPa), as functions of the strain
amplitude, $\gamma_0$, and the potential energy after 4000 loading
cycles, $U$. The horizontal dashed lines indicate $G$ and $\sigma_Y$
for the steadily sheared glass before cyclic loading was applied.
The straight red line in the panel (b) is the best fit to the data.}
\label{fig:G_sigY_vs_gamma0_U_amp_020_055}
\end{figure}

We next examine the stress-strain response of the binary glass
subjected to startup continuous shear deformation with a constant
strain rate of $10^{-5}\,\text{ps}^{-1}$. The dependence of shear
stress as a function of strain is shown in
Fig.\,\ref{fig:stress_strain_const_amps_upto_055} after loading for
4000 cycles at the indicated strain amplitudes. For comparison, the
data for the glass after rapid cooling but before cyclic deformation
are also included in
Fig.\,\ref{fig:stress_strain_const_amps_upto_055}. As expected, the
rapidly cooled glass under steady strain exhibits a smooth crossover
to plastic flow (the violet curve in
Fig.\,\ref{fig:stress_strain_const_amps_upto_055}). By contrast,
plastic flow of mechanically annealed glasses is preceded by the
yielding peak, which becomes more pronounced in glasses previously
loaded at larger strain amplitudes. In
Fig.\,\ref{fig:G_sigY_vs_gamma0_U_amp_020_055}, we summarize results
for the shear modulus, computed from the slope of stress-strain
curves at $\gamma_{xz} \leqslant 0.01$, and the peak value of the
stress overshoot as functions of the strain amplitude and the
potential energy after 4000 cycles. It can be concluded from
Fig.\,\ref{fig:G_sigY_vs_gamma0_U_amp_020_055}\,(b) that the
potential energy of the mechanically annealed glass is approximately
linearly related to the magnitude of the yielding peak. On the other
hand, the shear modulus remains rather insensitive to the loading
amplitude, as shown
Fig.\,\ref{fig:G_sigY_vs_gamma0_U_amp_020_055}\,(c).

\vskip 0.05in

\begin{figure}[t]
\includegraphics[width=12.0cm,angle=0]{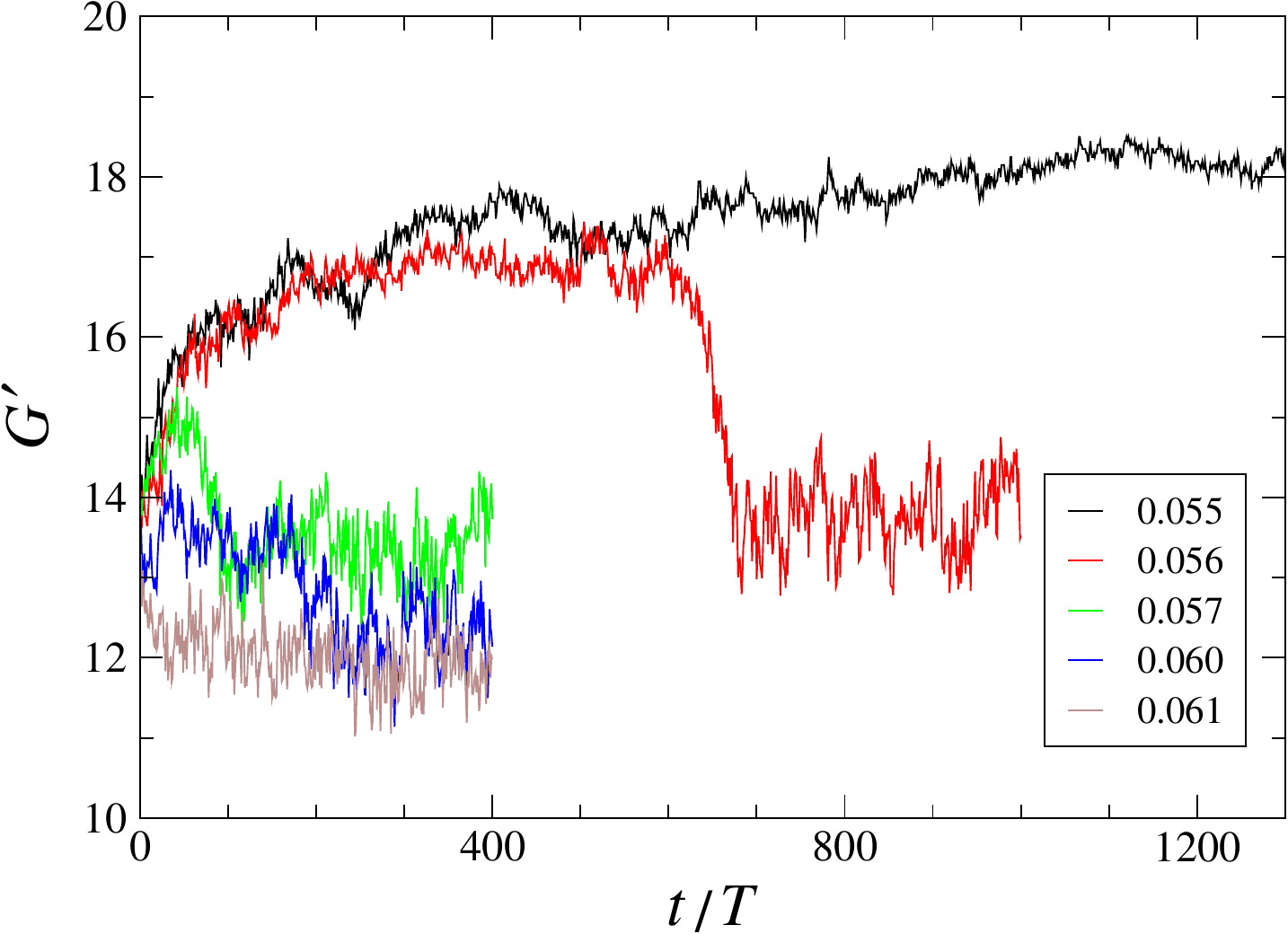}
\caption{The storage modulus $G^{\prime}$ (in units of GPa) versus
cycle number for strain amplitudes $\gamma_0 \geqslant 0.055$. The
data for $G^{\prime}$ at $\gamma_0 = 0.055$ are the same as in
Fig.\,\ref{fig:Gp_4000_T300_r10e12_amp_020_055}. The oscillation
period is $T=1.0\,\text{ns}$.}
\label{fig:Gp_4000_T300_r10e12_amp_055_56_57_60_61}
\end{figure}

The storage modulus as a function of the cycle number is plotted in
Fig.\,\ref{fig:Gp_4000_T300_r10e12_amp_055_56_57_60_61} for
$\gamma_0 \geqslant 0.055$. Note that the data for the strain
amplitude $\gamma_0 = 0.055$ are the same as in
Fig.\,\ref{fig:Gp_4000_T300_r10e12_amp_020_055}. It is seen in
Fig.\,\ref{fig:Gp_4000_T300_r10e12_amp_055_56_57_60_61} that in the
range $0.056 \leqslant \gamma_0 \leqslant 0.060$, cyclic loading
leads to an increase in $G^{\prime}$ for a number of transient
cycles, followed by the yielding transition, which is marked by an
abrupt drop in storage modulus.  Interestingly, the storage modulus
as a function of the cycle number at the critical strain amplitude,
$\gamma_0 = 0.056$, closely follows the data for $\gamma_0 = 0.055$
during the first 600 cycles. Thus, the onset of yielding cannot be
determined from the stress variation during the initial stage of
deformation near the critical strain amplitude. As shown in
Fig.\,\ref{fig:Gp_4000_T300_r10e12_amp_055_56_57_60_61}, the number
of transient cycles is reduced at higher strain amplitudes,
$\gamma_0 > 0.056$. Overall, these results correlate well with
dependence of the potential energy minima on the number of cycles
reported in Fig.\,\ref{fig:poten_T300_r10e12_amp_055_56_57_60_61}.

\vskip 0.05in


At the microscopic level, plastic deformation in disordered
materials can be described via the so-called nonaffine displacements
of atoms with respect to their neighbors~\cite{Falk98}. In practice,
the nonaffine quantity for an atom displaced from the position
vector $\mathbf{r}_{i}(t)$ to $\mathbf{r}_{i}(t+\Delta t)$ during
the time interval $\Delta t$ is computed by minimizing the following
expression:
\begin{equation}
D^2(t, \Delta t)=\frac{1}{N_i}\sum_{j=1}^{N_i}\Big\{
\mathbf{r}_{j}(t+\Delta t)-\mathbf{r}_{i}(t+\Delta t)-\mathbf{J}_i
\big[ \mathbf{r}_{j}(t) - \mathbf{r}_{i}(t)  \big] \Big\}^2,
\label{Eq:D2min}
\end{equation}
where $\mathbf{J}_i$ is the transformation matrix and the sum is
taken over neighboring atoms that are initially located within
$4.0\,\text{\AA}$ from $\mathbf{r}_{i}(t)$. It was recently shown
that in glasses under periodic deformation, rearrangements of atoms
become irreversible when their nonaffine displacements are greater
than a cage size~\cite{Priezjev16}. In addition, previous MD
simulation study has indicated that the cage size is about
$0.6\,\text{\AA}$ for $\text{Cu}_{50}\text{Zr}_{50}$ metallic alloy
near the glass transition temperature~\cite{Douglas19}.

\vskip 0.05in

\begin{figure}[t]
\includegraphics[width=12.0cm,angle=0]{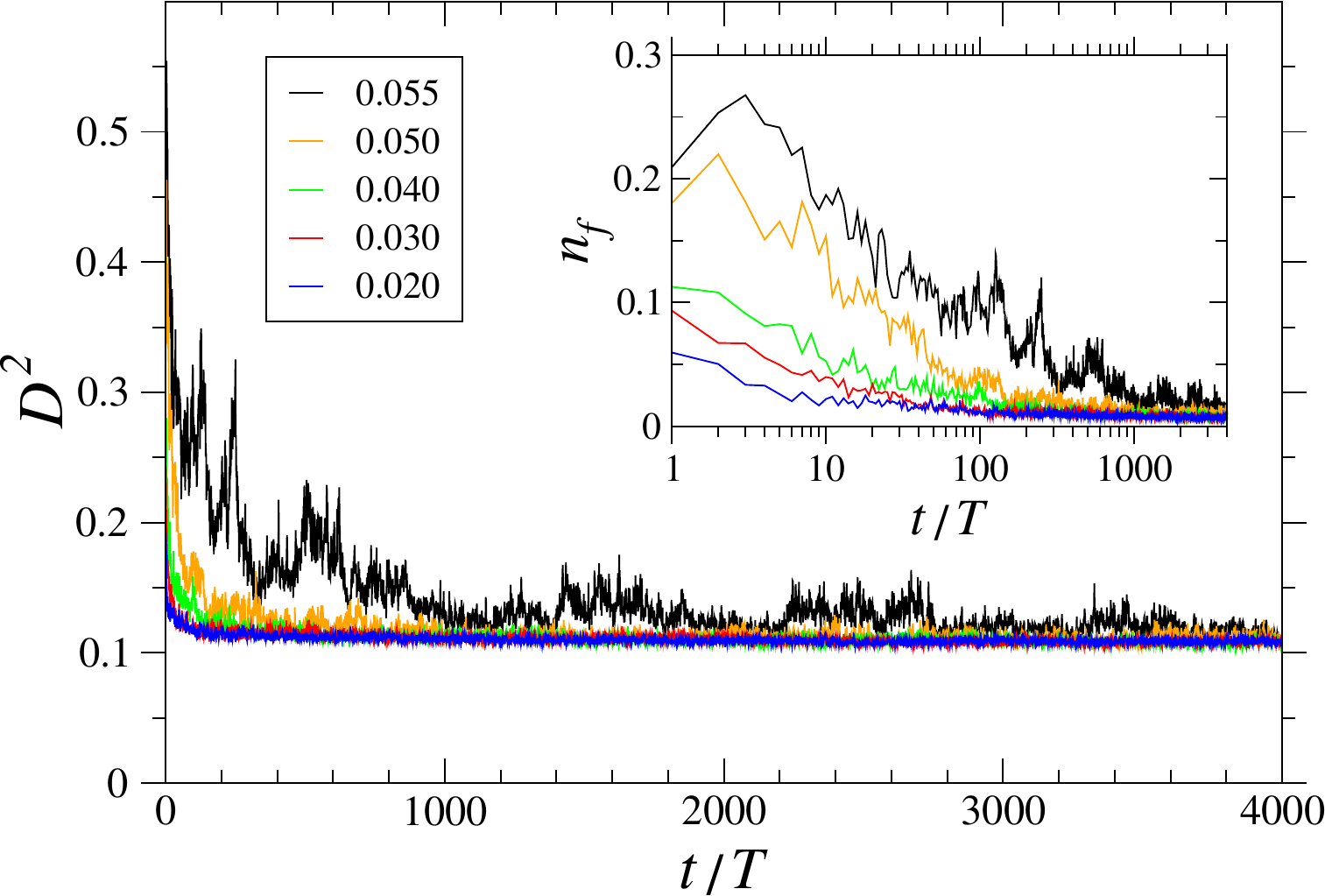}
\caption{The average of the nonaffine quantity, $D^2[(n-1)\,T,T]$
(in units of $\text{\AA}^2$), as a function of the number of cycles
for the indicated strain amplitudes. The oscillation period is
$T=1.0\,\text{ns}$. The inset shows the fraction of atoms with
$D^2[(n-1)\,T,T]>0.49\,\text{\AA}^2$ versus cycle number for the
same strain amplitudes. }
\label{fig:d2min_ave_ncyc_amp_020_055_inset_gt049}
\end{figure}
\begin{figure}[t]
\includegraphics[width=12.0cm,angle=0]{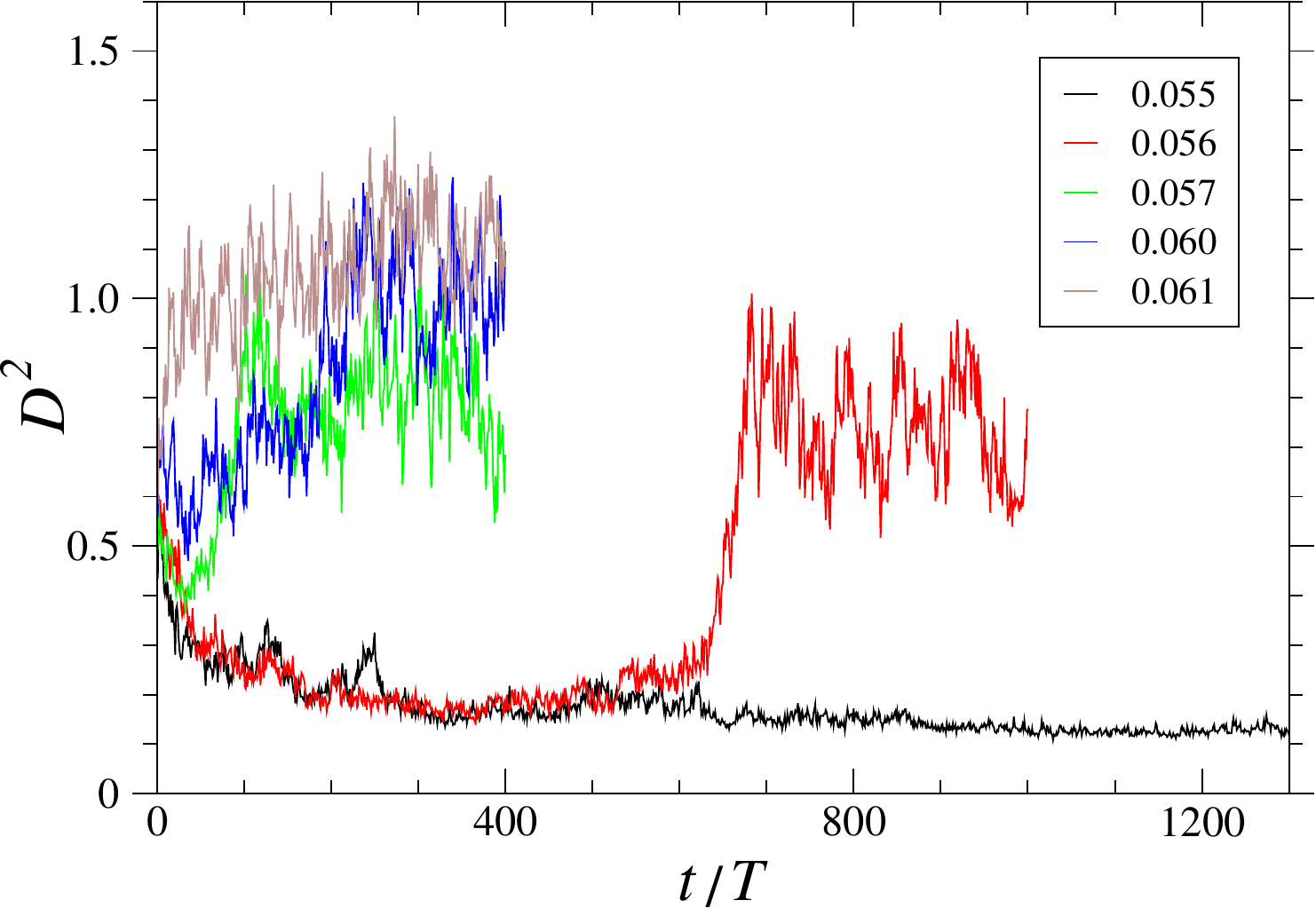}
\caption{The average of $D^2[(n-1)\,T,T]$ (in units of
$\text{\AA}^2$) versus number of cycles for strain amplitudes
$\gamma_0 \geqslant 0.055$. The data for the strain amplitude
$\gamma_0 = 0.055$ (the black curve) are the same as in
Fig.\,\ref{fig:d2min_ave_ncyc_amp_020_055_inset_gt049}. }
\label{fig:d2min_ave_ncyc_amp_055_56_57_60_61}
\end{figure}
\begin{figure}[t]
\includegraphics[width=12.0cm,angle=0]{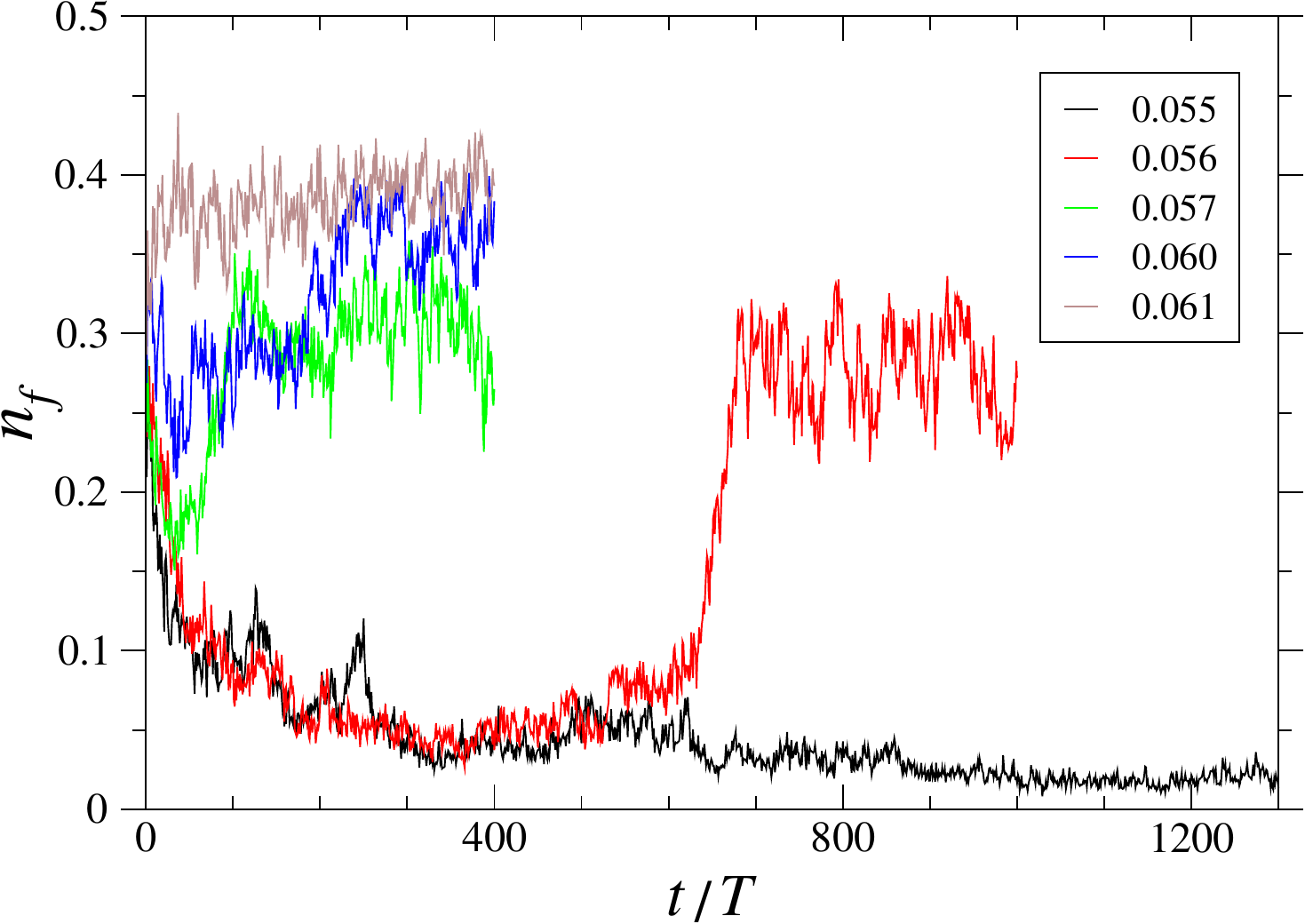}
\caption{The fraction of atoms with large nonaffine displacements,
$D^2[(n-1)\,T,T]>0.49\,\text{\AA}^2$, as a function of the number of
cycles for the tabulated strain amplitudes. The oscillation period
is $T=1.0\,\text{ns}$.}
\label{fig:nf_d2min_gt049_ncyc_amp_055_56_57_60_61}
\end{figure}

The dependence of the nonaffine measure as a function of the cycle
number is presented in
Fig.\,\ref{fig:d2min_ave_ncyc_amp_020_055_inset_gt049} for $\gamma_0
\leqslant 0.055$ and in
Fig.\,\ref{fig:d2min_ave_ncyc_amp_055_56_57_60_61} for $\gamma_0
\geqslant 0.055$.  In our analysis, the nonaffine measure was
computed at the beginning and end of each cycle, $\Delta t = T$ in
Eq.\,(\ref{Eq:D2min}), and then averaged over all atoms. For cyclic
loading below the critical strain amplitude, $\gamma_0 = 0.056$, the
glass evolves toward lower energy states via a sequence of plastic
events, whose size is reduced at lower amplitudes, and, as a result,
the average of $D^2$ is smaller at lower $\gamma_0$, as shown in
Fig.\,\ref{fig:d2min_ave_ncyc_amp_020_055_inset_gt049}. Note that
the largest fluctuations in $D^2$ occur at the strain amplitude
$\gamma_0 = 0.055$, whereas the lower bound, $D^2 \approx
0.1\,\text{\AA}^2$, is dominated by displacements of atoms within
their cages.  Further, the results in the inset to
Fig.\,\ref{fig:d2min_ave_ncyc_amp_020_055_inset_gt049} for the
fraction of atoms with large nonaffine displacements demonstrate
clearly that the number of atoms involved in plastic events decays
with increasing number of cycles. By contrast, the average of $D^2$
increases sharply at the yielding transition for $\gamma_0 \geqslant
0.056$ (see Fig.\,\ref{fig:d2min_ave_ncyc_amp_055_56_57_60_61}),
which is in agreement with the critical behavior of $U$ and
$G^{\prime}$ reported in
Figures\,\ref{fig:poten_T300_r10e12_amp_055_56_57_60_61} and
\ref{fig:Gp_4000_T300_r10e12_amp_055_56_57_60_61}, respectively.
Moreover, after the yielding transition, plastic flow is localized
within a shear band that contains a large fraction of atoms, $n_f
\gtrsim 0.25$ (see
Fig.\,\ref{fig:nf_d2min_gt049_ncyc_amp_055_56_57_60_61}).

\vskip 0.05in

\begin{figure}[t]
\includegraphics[width=12.0cm,angle=0]{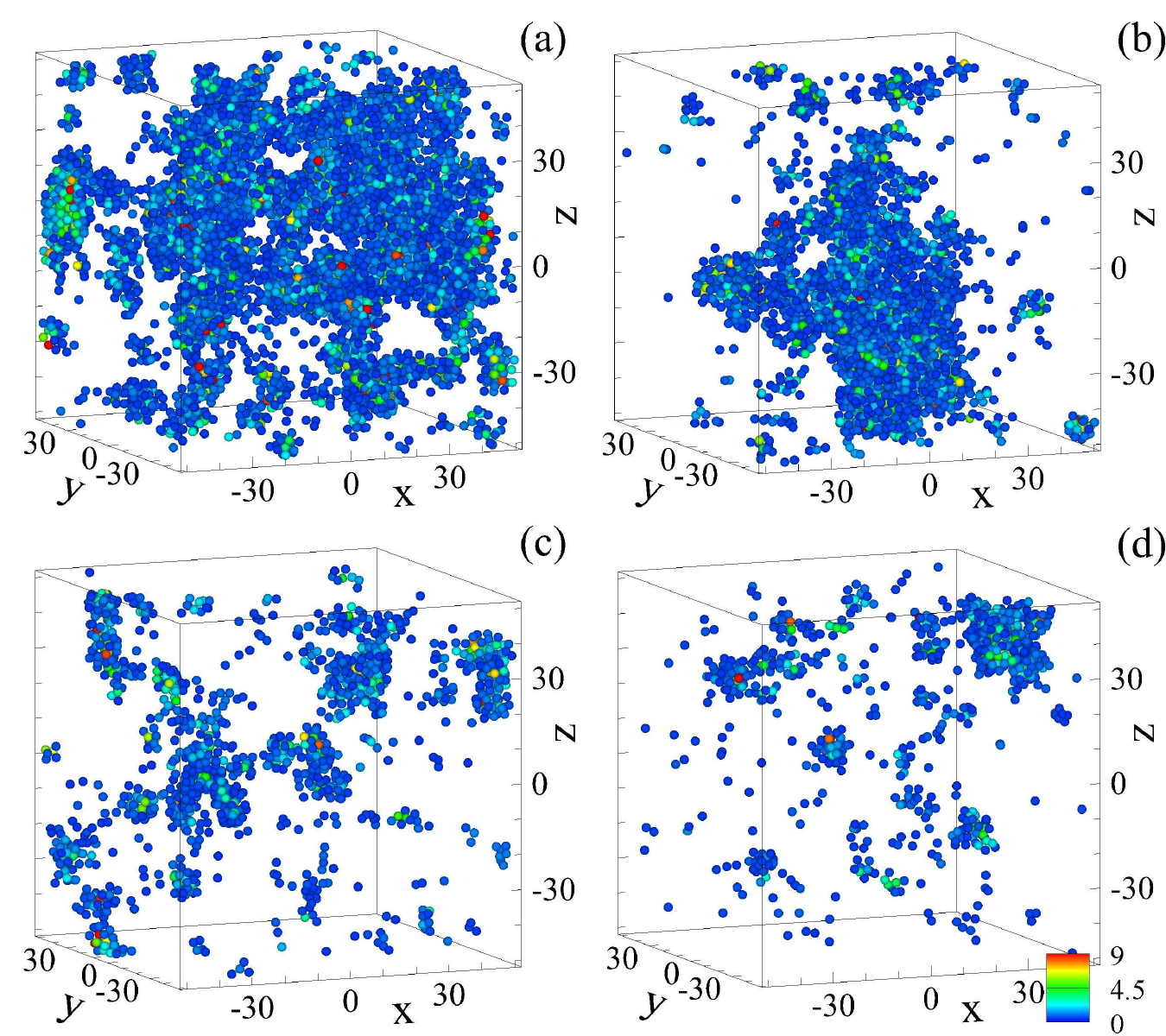}
\caption{Selected configurations of atoms in the
$\text{Cu}_{50}\text{Zr}_{50}$ glass loaded at the strain amplitude
$\gamma_0=0.055$. The nonaffine quantity is (a)
$D^2(50\,T,T)>0.49\,\text{\AA}^2$, (b)
$D^2(500\,T,T)>0.49\,\text{\AA}^2$, (c)
$D^2(1000\,T,T)>0.49\,\text{\AA}^2$, and (d)
$D^2(3000\,T,T)>0.49\,\text{\AA}^2$. Cu and Zr atoms are not drawn
to scale. The legend in the panel (d) indicates the magnitude of the
nonaffine quantity. }
\label{fig:snapshot_amp055}
\end{figure}
\begin{figure}[t]
\includegraphics[width=12.0cm,angle=0]{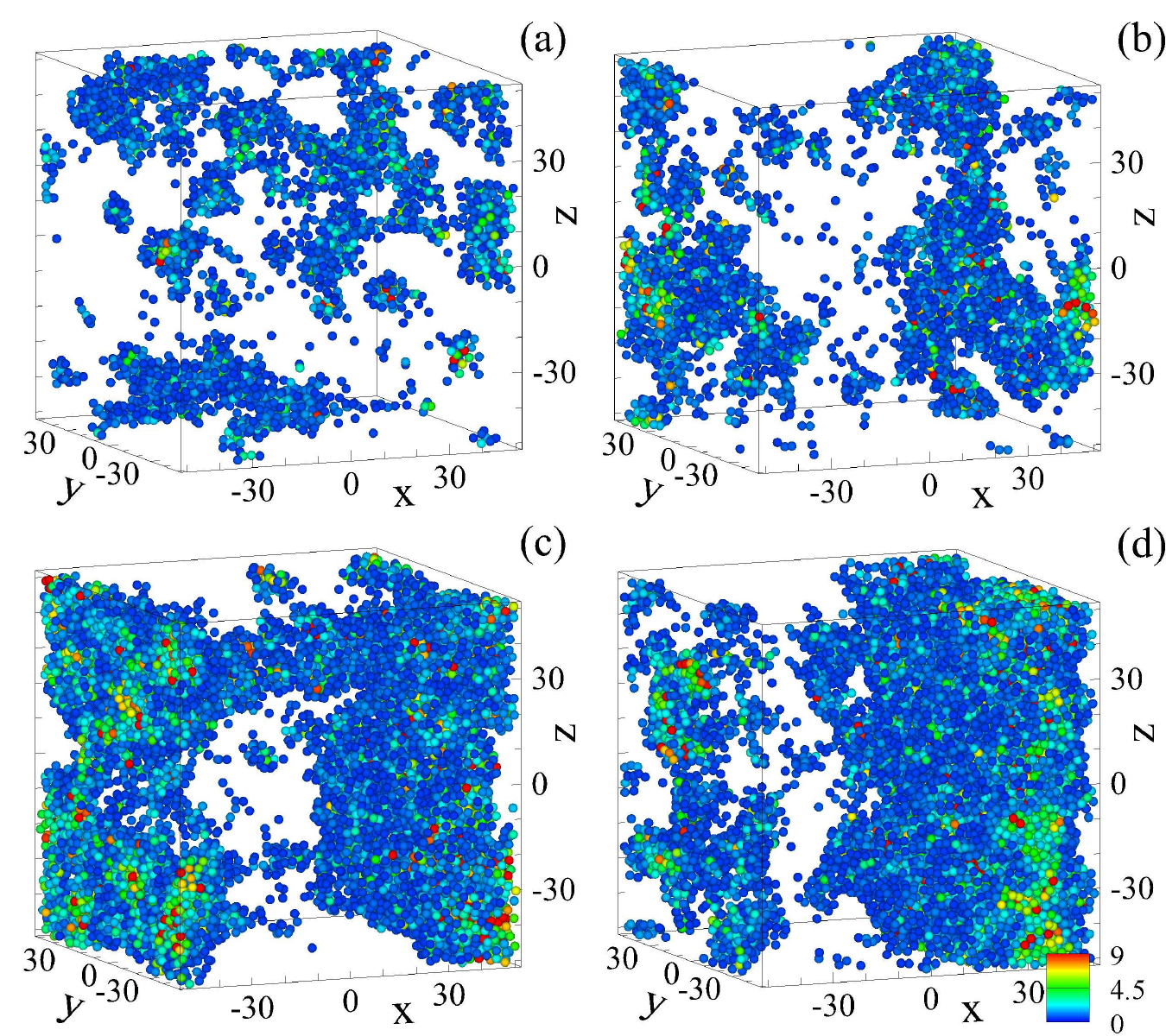}
\caption{Snapshots of atomic configurations of the binary glass
periodically deformed at the critical strain amplitude
$\gamma_0=0.056$. The nonaffine measure is (a)
$D^2(300\,T,T)>0.49\,\text{\AA}^2$, (b)
$D^2(600\,T,T)>0.49\,\text{\AA}^2$, (c)
$D^2(660\,T,T)>0.49\,\text{\AA}^2$, and (d)
$D^2(800\,T,T)>0.49\,\text{\AA}^2$. Atoms are not depicted to scale.
The colorcode is the same as in Fig.\,\ref{fig:snapshot_amp055}. }
\label{fig:snapshot_amp056}
\end{figure}

The spatial organization of plastically deformed domains during the
relaxation process at the strain amplitude $\gamma_0 = 0.055$ is
illustrated in Fig.\,\ref{fig:snapshot_amp055}.  The snapshots
include configurations of atoms after $n$ cycles at zero strain, and
the colorcode denotes the magnitude of the nonaffine measure for
$\Delta t = T$ in Eq.\,(\ref{Eq:D2min}). Note that atoms with
relatively small nonaffine displacements,
$D^2[(n-1)\,T,T]<0.49\,\text{\AA}^2$, were omitted for clarity, and,
therefore, the void space represents elastically deformed regions.
As shown in Fig.\,\ref{fig:snapshot_amp055}\,(a), the transition to
lower energy states initially proceeds via large irreversible
displacements of atoms that form several compact clusters
homogeneously distributed in the sample. Upon further loading, the
total number of atoms with large nonaffine displacements is
significantly reduced, see Fig.\,\ref{fig:snapshot_amp055}\,(b), and
eventually plastic rearrangements remain localized only in several
isolated clusters, see Fig.\,\ref{fig:snapshot_amp055}\,(c,\,d).
Hence, these results demonstrate that poorly annealed glasses
subjected to prolonged low-amplitude periodic deformation become
more stable and reversible.   On the contrary, cyclic loading at the
critical strain amplitude, $\gamma_0 = 0.056$, leads to a delayed
yielding transition and formation of a shear band across the whole
sample, as shown in Fig.\,\ref{fig:snapshot_amp056}. A close
comparison of the results for $\gamma_0 = 0.056$ in
Figs.\,\ref{fig:nf_d2min_gt049_ncyc_amp_055_56_57_60_61} and
\ref{fig:snapshot_amp056} reveals details of the yielding
transition; namely, shear band initiation [$n_f = 0.07$ in
Fig.\,\ref{fig:snapshot_amp056}\,(b)], propagation of plastic
regions along the $yz$ plane [$n_f = 0.20$ in
Fig.\,\ref{fig:snapshot_amp056}\,(c)], and formation of a fully
developed shear band [$n_f = 0.28$ in
Fig.\,\ref{fig:snapshot_amp056}\,(d)]. Together, these findings
emphasize the role of plastic rearrangements in structural
relaxation and yielding of metallic glasses under cyclic loading.

\section{Conclusions}

In summary, we investigated the critical behavior and mechanical
annealing of a Cu-Zr metallic glass subjected to periodic shear
deformation by means of molecular dynamics simulations. The glass
was prepared via rapid cooling from the liquid state to room
temperature and then subjected to oscillatory shear at strain
amplitudes ranged from slightly above to well below a critical
value. It was found that low-amplitude cyclic loading gradually
relocates the glass to lower energy states, and the storage modulus
approximately follows a logarithmic dependence on the number of
shear cycles. In addition, we showed that the potential energy after
four thousand shear cycles depends linearly on the yielding stress
of the annealed glass under monotonic shear deformation. The
structural relaxation proceeds via a sequence of plastic
rearrangements of clusters of atoms whose typical size is larger at
higher strain amplitudes. By contrast, when the strain amplitude is
greater than a critical value, the glass is mechanically annealed
for a number of cycles followed by an abrupt increase in the
potential energy due to flow localization within a shear band. The
formation of the shear band is accompanied with the increase of the
fraction of atoms with large nonaffine displacements and a drop in
storage modulus.

\section*{Acknowledgments}

The financial support from the National Science Foundation
(CNS-1531923) is gratefully acknowledged. MD simulations were
carried out at the Wright State University's Computing Facility and
the Ohio Supercomputer Center using the LAMMPS code~\cite{Lammps}.



%

%

%

%

%

%

%

%

%

%

%

%

\bibliographystyle{prsty}

\end{document}